\documentclass[twocolumn,preprintnumbers,superscriptaddress,amsmath,floatfix,amssymb,prd]{revtex4}
\usepackage{amsmath,amssymb,bm}
\usepackage[dvips]{graphicx}
\usepackage{hyperref}
\usepackage{epsfig}
\usepackage{amsfonts}
\usepackage{latexsym}
\usepackage{amsmath}
\usepackage{amssymb}
\usepackage{slashed}
\usepackage{longtable}
\usepackage{yfonts}
\usepackage{color}
\newcommand{\beq}{\begin{eqnarray}}
\newcommand{\eeq}{\end{eqnarray}}

\begin{document}
%
%
%
%

\title{Dissipative Charged Fluid in a Magnetic Field}

\author{Navid Abbasi}

\affiliation{School of Particles and Accelerators, Institute for Research in Fundamental Sciences (IPM), P.O. Box 19395-5531, Tehran, Iran}

\author{ Ali Davody}
\affiliation{School of Particles and Accelerators, Institute for Research in Fundamental Sciences (IPM), P.O. Box 19395-5531, Tehran, Iran}


\begin{abstract}
	{We study  the collective excitations  in a dissipative charged fluid at zero chemical potential when an external magnetic field is present. While in the absence of magnetic  field,  four collective excitations
	appear in the fluid, we find five hydrodynamic modes in presence of magnetic field. This implies that the magnetic field splits the degeneracy between the transverse shear modes. Using linear response theory, we then compute the retarded response functions. In particular, it turns out that the correlation between charge and the energy fluctuations will no longer vanish, even at zero chemical potential. By use of the response functions, we also derive the relevant Kubo formulas for the transport coefficients.}
\end{abstract}
\maketitle
\section{Introduction}\label{1}
Study of hydrodynamic limit in relativistic and non-relativistic systems has been an important issue since many years ago (See \cite{Schaefer:2014awa} and references therein.).  In particular, study of transport phenomena in an external field, has attracted considerable attention recently. As a remarkable progress,  it has been shown that in systems with chiral anomalies, the second law of thermodynamics necessitates the presence of parity odd terms in the constitutive relation of hydrodynamic current \cite{Son:2009tf}.

To investigate the dissipative character of fluids, one way is to study the collective excitations around equilibrium in the system. In this paper we will consider a charged fluid at zero chemical potential 
and compute the dispersion relation of the collective excitations in presence of an external magnetic field.
We show that the presence of magnetic field splits the degeneracy between the transverse shear modes. While in the absence of magnetic field there are four independent hydrodynamic modes, we find five long wave modes here. Of these five, two are longitudinal sound waves and three others are purely dissipative shear modes.  Moreover, compared to zero magnetic field limit, there appears new dissipative contributions in all hydrodynamic modes. In other words,  the presence of magnetic field intensifies the dissipation processes in a charged fluid. We will indicate it is actually the Lorentz force exerted from the magnetic field on the induced electric currents in the system that makes the fluid modes more quickly dissipate. 
 
 As known, the microscopic information of fluids are reflected by the 
 transport coefficients. These coefficients are in general related to equilibrium response functions. In order to determine how one could evaluate the transport coefficients in a magnetic field, we first compute the retarded response functions $G^R_{ab}$ in presence of a magnetic field. Then, from $SO(3)$ covariant response functions, we derive relevant Kubo formulas for conductivity, shear and bulk viscosity as well.  
 
 Since the presence of magnetic field breaks time reversal symmetry in the system, the retarded functions $G^R_{ab}$ are no longer symmetric under the exchange of "a" and "b" \cite{Kovtun:2012rj}. Specifically, the correlation function between charge and momentum density fluctuations, namely $G^R_{\pi_{i} n}$, which was already vanishing, turns out to be non-zero and more importantly, anti-symmetric. We physically explain why the magnetic field forces the fluctuations in charge density to be correlated with those in  momentum density. We also indicate, that the $G^R_{\pi_{i} n}$ is anti-symmetric,  is the consequence of this fact that in presence of magnetic field, the symmetry under time reversal holds if the sign of magnetic field is simultaneously reversed.
 
\section{Relativistic Fluid Dynamics}
In grand canonical ensemble at $\mu\ne 0$, the equilibrium state is characterized
by the density operator:
\begin{equation}
\hat{\rho}=\,\frac{1}{Z} e^{\beta u_{\mu}P^{\mu}+\beta \mu N}
\end{equation}
where $Z=\text{tr}e^{\beta u_{\mu}P^{\mu}+\beta \mu N}$ is the partition function. In the expression above, $P^{\mu}$ is the momentum operator, and $N$ is the conserved charge operator. This means that in a general Lorentz frame, we need to know five parameters to specify the equilibrium state of a system: a temperature $T$, a chemical potential $\mu$ and three components of four-velocity $u^{\mu}$ $(u^{\mu}u_{\mu}=-1)$.

 Hydrodynamics describes the dynamics of a thermal system slightly deviated from its thermal equilibrium,  or equivalently a locally equilibrated system, evolving towards its global equilibrium. To describe the hydrodynamic states, 
it is sensible to choose the slowly varying local functions $T(x)$, $\mu(x)$, $u^{\mu}(x)$ 
 as the hydrodynamic variables. From the hydrodynamic equations which are simply the conservation laws of the conserved global currents, one can find the solution for the hydrodynamic variables, just as a boundary value problem.
 
 In a general relativistic system with a $U(1)$ global symmetry, the global currents are the energy momentum tensor $T^{\mu \nu}$ and a current $J^{\mu}$.
 So in presence of an external background field the hydrodynamic equations take the form
\begin{equation}\label{2.3}
	\begin{split}
		\partial_{\mu}T^{\mu \nu}=& \,F^{\mu \nu}J_{\nu}\\
		\partial_{\mu}J^{\mu}=&\, 0
	\end{split}
\end{equation}
where the $F^{\mu \nu}J_{\nu}$ term in the right hand side of the first equation is the work which Lorentz force performs on the flow. Obviously, the number of unknown components ($10$ components of $T^{\mu \nu}$ and $4$ components of  $J^{\mu}$) are more than the number of equations (4+1)\footnote{We will always consider a system in $3+1$ dimension through this note. The generalization to higher spacial dimensions is straightforward.}. However, the idea of local thermal equilibrium is that one can express $T^{\mu\nu}$ and $J^{\mu}$
 as functions of exactly five slowly varying hydrodynamic variables.
 
 Near the local equilibrium also, $T^{\mu \nu}$ and $J^{\mu}$ may be expressed in derivative expansion of hydrodynamic fields. Up to first order in derivatives, the constitutive relations are given by:
 \begin{eqnarray}
  T^{\mu \nu}&=& \,(\epsilon+p) u^{\mu} u^{\nu}+ p \eta^{\mu \nu}+ \tau^{\mu\nu}\\
  J^{\mu}&=& \,n u^{\mu} +\nu^{\mu}.
 \end{eqnarray}
  In the expressions above, the dissipative parts, namely $\tau^{\mu\nu}$ and $\nu^{\mu}$, are not uniquely determined\footnote{There is an ambiguity with defining the fluid velocity out of equilibrium in relativistic fluid dynamics. Here we choose the Landau-lifshitz frame to fix this ambiguity and refer the interested reader to have a look at \cite{Bhattacharya:2011tra} for detailed explanations.}.
  In the so-called Landau-Lifshitz frame \cite{Landau} we have:
\begin{eqnarray}\label{disspart}
 \tau^{\mu \nu}=-\eta P^{\mu\alpha}P^{\nu\beta}\left(\partial_{\alpha}u_{\beta}+\partial_{\beta}u_{\alpha} \right)-\left(\zeta-\frac{2}{3}\eta\right) P^{\mu \nu} \partial.u\\
 \nu^{\mu}=  - \sigma T P^{\mu \nu} \partial_{\nu}\left(\frac{\mu}{T}\right)+\sigma E^{\mu}\,\,\,\,\,\,\,\,\,\,\,\,\,\,\,\,\,\,\,\,\,\,\,\,\,\,\,\,\,\,\,\,\,\,\,\,\,\,\,\,\,\,\,\,\,\,\,\,\,\,\,\,\,\,\,\,\,\,\,\,\,\,\,\,
\end{eqnarray}
  with $P^{\mu \nu}=u^{\mu} u^{\nu}+ \eta^{\mu \nu}$.
 Here, $\eta$, $\zeta$ and $\sigma$ are non-negative transport coefficients: shear viscosity, bulk viscosity and conductivity respectively. 
  We may also define the electric field in the rest frame of the fluid: $E_{\mu}\equiv F_{\mu \nu} u^{\nu}$.
  For  consistency with derivative counting in hydrodynamic,
 we take the strength of the background gauge field $A_{\mu}$
 of the same order of the  temperature and the chemical potential; so $A_{\mu}\sim O(\partial^0)$ and $F_{\mu\nu} \sim O(\partial)$ \cite{Son:2009tf}.
 This assumption constrains the value of magnetic field in our problem;   it turns out that for hydrodynamic perturbation with momentum $k$, the magnetic field must be of order of $B\sim \sqrt{\eta/\sigma} \,k$. Such magnetic field could affect the velocity of sound in magnetized fluid. Since our system  is assumed to be non-magnetized here, the effect of magnetic field appears at least from the first order in derivative expansion. We expect the magnetic field makes contributions of order of $\sigma \textbf{B}^2 \sim \eta \textbf{k}^2$ to dissipative processes.

 \section{Hydrodynamic Fluctuations At Zero Chemical Potential}
 In order to study the hydrodynamic fluctuations we have to determine the state of equilibrium and then linearize the equations around that. Before proceeding further, let us denote that in what follows, rather than choosing $T$, $\mu$ and $u^{i}$ as the hydrodynamic variables, we will choose their conjugate variables, the energy density $\epsilon$, momentum density $\pi^i$ and charge density $n$. The importance of this choice is that the latter have microscopic definitions given by operators $T^{00}(x)$, $T^{0i}(x)$ and $J^{0}(x)$.
  \subsection{State of Equilibrium in presence of an External Magnetic Field}
  Each equilibrium state of the system is specified by a static solution of the equations (\ref{2.3}).  It is straightforward to see that the set $\{u^{\mu}=(1,\textbf{0}),\, \mu=\text{const.}, \,T=\text{const.}\}$ not only characterizes state of equilibrium in the absence of external field, it can also
  be a thermodynamic solution when a constant background magnetic field is present.
  It is physically reasonable: magnetic field does not affect the static charges\footnote{The situation is different in the case of electric field; since the electric field exerts force even on the static charges, it has to be absent in equilibrium.}.
  
 The case what we are interested in is exactly the above-mentioned equilibrium state but at $\mu=0$.
Let us remind that a state in grand canonical ensemble at $\mu=0$ has no charge, i.e. $\bar{n}=0$\footnote{Through this note, a "bar" over every variable denotes the equilibrium value of that variable.}.  In relativistic hydrodynamics, this means that there is an equal number of particles and anti particles in thermal equilibrium. Hydrodynamic fluctuations in this state however would be a sensible thing, namely one can compute $\delta\mu$ and $\delta n$ around equilibrium\footnote{In contrast, in non-relativistic systems, $\bar{n}=0$ means that the equilibrium state is purely neutral, with no charge that 
could flow.}.

In summary, our goal is to compute the correlation between hydrodynamic fields around the equilibrium state
\begin{equation*}\label{2.5}
 \bar{u}^{\mu}=(1,\textbf{0}),\,\,\bar{T}=\text{const.},\,\,\bar{\mu}=0,\,\,\textbf{E}=\textbf{0}, \,\,\textbf{B}=\text{\textbf{const.}}.
  \end{equation*}
 
  \subsection{Hydrodynamic Modes in Presence of Magnetic Field}
Having specified the state of equilibrium, we have to linearize equations around that. Let us remind that in  the absence of magnetic field at $\bar{\mu}=0$, the fluctuations in charge density decouple from the fluctuations of momentum and energy density (see (\ref{disspart}). However when $\textbf{B}$ is present in equilibrium, even at $\bar{\mu}=0$,
the fluctuating fields $n(t, \textbf{k})$, $\frac{}{}\pi_i(t, \textbf{k})$ and $\delta\epsilon(t, \textbf{k})$ are coupled to each other as the following:  
\begin{equation}\label{lineareq}
\begin{split}
\partial_{t}  n+ \,\textbf{k}^2 D  n +\frac{i  \sigma}{\bar{w}}  k_{j} F^{j k} \pi_{k}&=0\\
\partial_{t}\pi^{j}+\,i k_{j} v_s^2 \delta \epsilon   - \mathcal{M}^{i j } \pi_{i}&=\,i D F^{j m} k_{m}  n+\,\frac{\sigma} {\bar{w}}F^{j k} F_{k m}\pi^{m}\\
\partial_{t} \delta \epsilon +\, i k_{j} \pi^{j}&=0\\
\end{split}
\end{equation}
 with $\mathcal{M}_{i j }=-\gamma_{\eta}(\textbf{k}^2 \delta_{i j}- k_{i} k_{j})- \gamma_{s} k_{i}k_{j}$, $\bar{w}=\bar{\epsilon}+\bar{p}$, $\gamma_{\eta}=\eta/\bar{w}$, $\gamma_{s}=\left(\frac{4}{3}\eta+\zeta \right)/\bar{w}$, $D=\sigma/\chi$ and $\chi=(\partial n/\partial\mu)_{\mu=0}$. The velocity of sound is defined through $v_s^2=\partial p/\partial\epsilon$
\footnote{Note that in writing equations \ref{lineareq}, we have used the spacial Fourier
transformed fields defining as $
\phi_{a}(t,\textbf{k})=\int d^3 \textbf{x} e^{i \textbf{k}.\textbf{x}} \phi_{a}(t, \textbf{x})$.}.

Equations (\ref{lineareq}) can also be written in a concise form. To proceed let us define 
the supefield $\phi_a(t,\textbf{k})=\left(n,\frac{}{} \pi^i,\frac{}{} \epsilon\right)$. In terms of $\phi_a$, the linearized equations can be written as 
\begin{equation}
\partial_{t} \phi_{a}(t, \textbf{k})+\,M_{a b}(\textbf{k})\, \phi_{b}(t, \textbf{k})=0
\end{equation}  
where the matrix $M$ is 
\begin{equation}
M_{a b}=
\left( {\begin{array}{ccc}
	\textbf{k}^2 D  & \frac{i \sigma}{\bar{w}} k^{m}F_{m j} & 0\\ 
	-i D F^{i m}k_m& -\mathcal{M}^{i}_{ j}-\frac{\sigma}{\bar{w}}F_{jm}F^{mi} &i k^{i} v_{s}^2  \\
	0    & i k_j & 0
	\end{array} } \right).
\end{equation}
Now, hydrodynamic modes may be simply found via solving the following equation:
\begin{equation}
\det\left(-i \omega\delta_{ab}+\frac{}{} M_{ab}(\textbf{k})\right)=0
\end{equation}
Doing so, we obtain five collective modes around the equilibrium (See table (\ref{tabelmodes})).
There are some interesting points regarding with these modes which we point them out in what follows. First, when $\textbf{B}=0$, the modes corresponding to the momentum fluctuation in the directions transverse to $\textbf{k}$ are degenerate, because the only preferred direction in the system is identified by the wave vector $\textbf{k}$. However, as expressed in table \ref{tabelmodes}\footnote{$
	a= \big(\sigma \textbf{B}^2/\bar{w}+(\gamma_s-\gamma_{\eta}\big)\textbf{k}^2)$,\,\newline
	 $b=\textbf{k}^2(\beta_1-\gamma_{\eta} \textbf{k}^2+\frac{\gamma_{s}\sigma}{\bar{w}}\textbf{B}^2\,\cos^2 \theta+\frac{\gamma_{\eta}\sigma}{\bar{w}}\textbf{B}^2\,\sin^2 \theta+\,\gamma_{s}\gamma_{\eta}\textbf{k}^2)$ and 	\newline
	 $c=-\beta_1 \textbf{k}^2 \big(\frac{\sigma}{\bar{w}}\textbf{B}^2\,\cos^2 \theta+\,\gamma_{\eta}\textbf{k}^2\big)$.		
}
, in presence of  magnetic field,  the degenerate modes, namely  $\omega_4$ and $\omega_5$,  are split into two dissipative modes.  Moreover, when $\textbf{B}\parallel\textbf{k}$, the degeneracy is restored again
\begin{equation}
\omega_{4}(\textbf{k})=\,\omega_{5}(\textbf{k})=\,-i\left(\gamma_{\eta}\textbf{k}^2+\frac{\sigma}{\bar{w}}\textbf{B}^2\right).
\end{equation}  
The point which has to be physically explained is how the magnetic field participates in dissipation processes. Let us briefly explain how it happens for the case of shear modes. Consider a spacial frame wherein $\textbf{k}=(k,0,0)$ and $\textbf{B}=(B_x,B_y,0)$. In this frame, the shear mode $\omega_5$ reflects the decay rate of  the momentum fluctuations in the  direction $y$. So we consider the momentum fluctuation as being $\pi_y$.  According to (\ref{disspart}), a momentum fluctuation in an external magnetic field, induces an electric current: 
\begin{equation}
J_z=-\frac{\sigma}{\bar{w}} B_{x} \pi_{y}.
\end{equation}
The magnetic field itself, exerts a Lorentz force on the current or equivalently on the unit volume of the fluctuating element of the fluid as being
\begin{equation}
d\textbf{F}=\,\textbf{J}\times \textbf{B}=\,\left(0,-\frac{\sigma}{\bar{w}} B_{x}^2 \pi_{y},0\right).
\end{equation}
No matter how the profile of $\pi_y$ is, the above force is always directed opposite to $\pi_y$.  
Therefore, the Lorentz force makes the momentum of fluid element dissipates and the rate of dissipation is exactly the same as magnetic contribution in $\omega_5$. 
With similar arguments, one can physically interpret the contribution of magnetic field to dissipative parts of the other hydrodynamic modes. 

As the second point, let us give a comment on the velocity of sound modes given in table \ref{tabelmodes}.
%
As it can be clearly seen, 
the velocity of sound in a charged fluid is not affected by the presence of a background magnetic field. This is not surprising since the fluid is supposed to be non-magnetized here.  If the fluid were magnetized, the pressure would not be isotropic even in equilibrium and therefore the sound velocity would be dependent on the direction of magnetic field. Investigating such effects would be of serious importance when studying magnetohydrodynamics in a charged system in presence of a strong magnetic field \cite{Huang:2011dc,Jensen:2011xb}. 
\newline

\begin{longtable*}{|l|c|}
	\hline
\,\,\,\,\,\,\,\,\,\,\,\,\,\,\,\,\,\,\,\,\,\,\,\,\,\,\,\,\,\,\,\,\,	Hydrodynamic modes in presence of magnetic field & Hydrodynamic modes at $\textbf{B}=0$   \\
	\hline
	& '\\
	$\omega_{1,2}(\textbf{k})=\,\pm v_{s} k-\,\frac{i}{2}\left(\textbf{k}^2\gamma_{s}+\,\frac{\sigma}{ 
		\bar{w}}\textbf{B}^2 \sin^2 \theta \right)$ &  	$\omega_{1,2}(\textbf{k})=\,\pm v_{s} k-\,\frac{i}{2}\textbf{k}^2\gamma_{s}$  \\
	& \\
	$	\omega_{3,4}(\textbf{k})= -\frac{i}{2}\left(\textbf{k}^2 (D+\gamma_{\eta})+ \frac{\sigma}{\bar{w}}\,\textbf{B}^2 \pm\sqrt{(\textbf{k}^2(D-\gamma_{\eta})- \frac{\sigma}{\bar{w}}\,\textbf{B}^2)^2+ \frac{4 D \sigma}{\bar{w}}\textbf{B}^2\,\textbf{k}^2\sin^2 \theta}\right)$ &	$\omega_{3}(\textbf{k})=-iD \textbf{k}^2$ ,\\
	& $\omega_{4}(\textbf{k})=-i\gamma_{\eta} \textbf{k}^2$\\
	$\omega_{5}(\textbf{k})=-i\left(\textbf{k}^2 \gamma_{\eta}+\frac{\sigma}{ \bar{w}}\textbf{B}^2 \cos^2 \theta\right)$ & $\omega_{5}(\textbf{k})=-i\gamma_{\eta} \textbf{k}^2$\\
	 & \\
	\hline
	\caption{Comparison between hydrodynamic modes "in presence" and "in absence" of external magnetic field at $\bar{\mu}=0$. In the expressions above, $\theta$ is the angle between momentum vector $\textbf{k}$ and the magnetic field $\textbf{B}$, i.e. $\cos\theta= \hat{k}.\hat{\textbf{B}}$.
		}\label{tabelmodes}
\end{longtable*}
\subsection{Response functions}
Using the methods of linear response theory, the response functions have been extensively discussed in non-relativistic hydrodynamics in \cite{Kadanoff}. Recently, Kovtun  applied the methods of linear response theory to the case of relativistic hydrodynamics. Specifically, in \cite{Kovtun:2012rj}, it has been shown that the retarded function
\begin{equation}
G^R_{a b}(t-t',\textbf{x}-\textbf{x}')= -i \theta(t-t') \left\langle [\hat{\phi}_a(t, \textbf{x}),\hat{\phi}_b(t',\textbf{x}')]\right\rangle 
\end{equation}
might be computed in the framework of linear response theory via using the following formula
\begin{equation}\label{retarded}
G^{R}_{a b}(\omega, \textbf{k})=-\left(\delta_{a c}+\frac{}{} i \omega (K^{-1})_{a c}\right) \chi_{c b}.
\end{equation}
In the expression above, $K_{ab}= - i \omega \delta_{a b}+ M_{a b}(\textbf{k})$ and  $\chi_{a b}$ is the static thermodynamic susceptibility which is defined by
\begin{equation}\label{suscep}
\chi_{ab}=\left(\frac{\partial \phi_a}{\partial \lambda_b}\right).
\end{equation}
Here $ \lambda_{a}$ is the external source term which couples to field $\phi_a$ through the perturbation Hamiltonian. In order to identify sources, we have to find the linear change in Hamiltonian, namely $\delta H(t, \textbf{x})$, for infinitesimal slowly varying disturbances $\delta T(t, \textbf{x})$, $\delta \mu(t, \textbf{x})$ and $ v^i(t, \textbf{x})$. To proceed, we may formally write
\begin{equation}\label{densitylocal}
\hat{\rho}=\frac{1}{Z}e^{\beta(x) \left(u_{\mu}(x)P^{\mu}(x)+\frac{}{}\mu(x)N(x)\right)}=\frac{1}{Z}e^{-\bar{\beta} \left(H+\frac{}{}\delta H(x)\right)}.
\end{equation}
Simplifying equation (\ref{densitylocal}) yields:
\begin{equation}
\begin{split}
&\delta H(x)=-\int d^3\textbf{x}\, \lambda_{a}(x) \hat{\phi}_a(x) =\\
&-\int d^3\textbf{x}\left(\frac{\delta T(x)}{ \bar{T}}T^{00}(x)
+ \delta\mu (x) J^0(x)+ v_{i}(x)T^{0i}(x) \right)
\end{split}
\end{equation}
Therefore we identify $\lambda_{a}=\left(\delta \mu\,,\, v_{i}\,,\,\frac{\delta T}{\bar{T}} \right)$ as the sources  corresponding to the hydrodynamic fields $\phi_a=\,(\delta n, \pi_i, \delta \epsilon)$. As a result,
 the susceptibility matrix (\ref{suscep}) is
 
\begin{equation}
\chi_{a b}=
\left( {\begin{array}{ccc}
T\left(\frac{\partial n}{\partial T}\right)_{\mu/T}     & 0  & \left(\frac{\partial n}{\partial \mu}\right)_{T}\\
0 & \bar{w} & 0\\
T\left(\frac{\partial \epsilon}{\partial T}\right)_{\mu/T}    & 0& \left(\frac{\partial \epsilon}{\partial \mu}\right)_{T}\end{array} } \right)\, =\,\left( {\begin{array}{ccc}
	\chi     & 0  & 0\\
	0 & \bar{w} & 0\\
	0    & 0 & \frac{\bar{w}}{c_s^2} 
\end{array} } \right)\, 
\end{equation}
Similar to the case where the magnetic field is absent \cite{Kovtun:2012rj}, the susceptibility matrix is diagonal here. This is due to this reason that chemical potential in our 
system is zero in equilibrium\cite{Kovtun:2012rj}.  

Now we substitute (\ref{suscep}) into formula (\ref{retarded}) to compute the retarded functions in presence of a constant external magnetic field. Results have been reported in table (\ref{tabelGR}), organized in terms of $SO(3)$ group representations.  In the scalar sector there are two non-vanishing response function. In oreder to represent the vector type response functions, we first have to find a vector basis. Using $\textbf{k}$ and $\textbf{B}$, one can consider three independent vectors
\begin{equation}\label{basis vectors}
\textbf{V}^{1}=\textbf{k},\,\,\,\,\textbf{V}^{2}=\cos\theta\,\textbf{B},\,\,\,\,\textbf{V}^{3}=\textbf{B}\times \textbf{k}
\end{equation}
where $\cos \theta= \hat{\textbf{B}}.\hat{\textbf{k}}$. Taking these three vectors as the basis, we have covariantly expressed the vector retarded functions, namely $G^R_{\pi_i \epsilon}$ and $G^R_{\pi_i n}$, in the second row of table (\ref{tabelGR}).

 The basis of symmetric tensors
may be constructed by using $\delta_{i j}$ plus all symmetric bi-vectors made out of basis vectors given above. It simply turns out that there are six independent symmetric tensors as being 
\begin{equation}
\delta_{i j},\,\,\,\,k_i k_j,\,\,\,\,\,\,B_i B_j,\,\,\,\,\,\,V^m_{(i}V^n_{j)} ,\,\,\,m\ne n;
\end{equation}
however only four of the them have participated in the
expression of $G^R_{\pi_i \pi_j}$.  
\begin{longtable*}{|l|c|c|}
	\hline
	\,\,\,$SO(3)$ Type & Retarded Green's Functions  \\
	\hline
	& '\\
	\,\,\,\,\,\,$\text{scalar}$ & $G^{R}_{\epsilon n}(\omega,\textbf{k})=0\,\,\,\,\,\,\,\,,G^{R}_{\epsilon \epsilon}(\omega,\textbf{k})=\,\frac{\bar{w}\, \textbf{k}^2 \Delta_{5}}{\Delta_{125}},\,\,\,\,\,\,\,\,
	G^{R}_{n n}(\omega,\textbf{k})=\,\frac{-i\sigma \, \textbf{k}^2 \Delta_{5}}{\Delta_{34}}
	$   \\
	&  \\
	\hline
	& '\\
	\,\,\,\,\,\,$\text{vector}$ & $G^R_{ \pi_i n}(\omega,\textbf{k})=\,-G^R_{n \pi_{i}}(\omega,\textbf{k})=\,\frac{\sigma \omega}{\Delta_{34}}\left(\textbf{B}\times\frac{}{}\textbf{k}\right)_i$\,\,\,\,\,\,\,\,\,\,\,\,\,\,\,\,\,\,\,\,\,\,\,\,\,\,\,\,\,\,\,\,\,\,\,\,\,\,\,\,\,\,\,\,\,\,\,\,\,\,\,\,\,\,\,\\
	& $G^R_{\epsilon \pi_i}(\omega,\textbf{k})=\,G^R_{\pi_{i}\epsilon}(\omega,\textbf{k})
	=\,\frac{\omega}{\Delta_{125}}\left(\bar{w}(\omega+i \textbf{k}^2 \gamma_{\eta})\,k_{i}+\frac{}{}i \sigma(\textbf{B}.\textbf{k}) B_{i}\right)$ \\
	& \\
	\hline
	& \\
	\,\,\,\,\,\,Tensor & $G^R_{\pi_i \pi_j }(\omega, \textbf{k})=\left(-\frac{i \bar{w} A_c}{\Delta_{5}}-\frac{i \sigma \omega^2}{\Delta_{345}}\frac{(\textbf{B}\times\textbf{k})^2}{\textbf{k}^2}\right)\delta_{ij}+\left(-\frac{\bar{w} \omega	(i a \omega^2 -b \omega +i c)}{\Delta_{1255}}+\frac{i \sigma \omega^2}{\Delta_{345}}\textbf{B}^2\right)\frac{k_i k_j}{\textbf{k}^2}$\\
	&\,\,\,\,\,\,\,\,\,\,\,\,\,\,\,\,\,\,\,\,\,\,\,\,\,\,\,\,\,\,\,\,\,\,\,\,\,\,\,\,\,\,\,\,\,\,\,\,\,\,\,\,\,\,\,\,\,\,\,\,\,\,\,\,\,$+\left(- \frac{\sigma^2 \omega^2}{\bar{w}\Delta_{1255}}\frac{(\textbf{B}.\textbf{k})^2}{\textbf{k}^2} +\frac{i\sigma \omega^2}{\Delta_{345}}\right)B_i B_j+\left(\frac{i \sigma \omega^2(\omega+i  \gamma_{\eta} \textbf{k}^2)}{\Delta_{1255}}-\frac{i \sigma \omega^2}{\Delta_{345}}\right)\left(\frac{\textbf{B}.\textbf{k}}{\textbf{k}^2}\right)k_{(i} B_{j)}$   \\
	& \\
	\hline
	\caption{Returded functions organized in different sectors of $SO(3)$ group. In above expressions $A_c=\frac{\sigma}{\bar{w}}\textbf{B}^2\,\cos^2 \theta+\,\gamma_{\eta}\textbf{k}^2 
		$ and by the $\Delta_{ij...}$ in denominators, we mean $(\omega-\omega_{i}(\textbf{k}))(\omega-\omega_{j}(\textbf{k}))\dots.$}\label{tabelGR}
\end{longtable*}
In the absence of magnetic field, retarded response functions $G^R_{ab}$ are symmetric under the exchange of a and b fields
\begin{equation}\label{Gab=Gba}
G^R_{ab}(\omega,\textbf{k})=G^R_{ba}(\omega,\textbf{k}).
\end{equation} 
The origin of the above symmetry is that in a system which is microscopically 
time-reversal invariant, the space-time translational invariance implies that\cite{Kovtun:2012rj}
\begin{equation}\label{Gab=etaGba}
G^R_{ab}(\omega,\textbf{k})=\eta_a \eta_b G^R_{ba}(\omega,-\textbf{k})
\end{equation} 
with $\eta_a$ being the time-reversal eigenvalue of the hydrodynamic field $\phi_a$. It is straightforward to show that when $\textbf{B}=0$, 
(\ref{Gab=etaGba}) leads to (\ref{Gab=Gba}).\footnote{For example:\newline
$G^R_{\pi_i \epsilon}(\omega,\textbf{k})=(-1)(+1)G^R_{\epsilon \pi_i}(\omega,-\textbf{k})=\,G^R_{\epsilon \pi_i}(\omega,\textbf{k})$\newline
where in the second equality we have used the point that
$G^R_{\epsilon \pi_i}(\omega,-\textbf{k})$ as a vector has to be made out of
$k_{i}$, so it must be in the form $G^R_{\pi_i \epsilon }(\omega,-\textbf{k})=(\text{scalar})k_i$.} 

 When $\textbf{B}$  is non-zero,  $\textbf{V}^2$ and $\textbf{V}^3$
may participate in the structure of vector response functions too.
On the other hand, the relation (\ref{Gab=etaGba}) has to be generalized in presence of magnetic field\footnote{Equation \ref{Gab=etaGbaB} is the basis for the Onsager relations \cite{Onsager1,Onsager2}.}
\begin{equation}\label{Gab=etaGbaB}
G^R_{ab}(\omega,\textbf{k},\textbf{B})=\eta_a \eta_b G^R_{ba}(\omega,-\textbf{k},-\textbf{B}).
\end{equation} 
Denoting these considerations, it turns out that  only $\textbf{V}^2$ is 
allowed to participate in the structure of $G^R_{\epsilon \pi_i}$ (See table (\ref{tabelGR}))\footnote{The behavior of $\textbf{V}^3$ under the transformations $\textbf{k}\rightarrow -\textbf{k}$ and $\textbf{B}\rightarrow -\textbf{B}$ is different from that of $\textbf{V}^1$ and $\textbf{V}^2$. Consequently, the basis vectors do not all  participate in the expression of a vector response function. If $\textbf{V}^3$ is present, $\textbf{V}^1$ and $\textbf{V}^2$ are absent and vice versa. }. 

The situation is somewhat different in the case of $G^R_{\pi_i n}$. Consider a momentum density fluctuation $\pi_x$ appeared at $t=0$ around the origin. In a magnetic field being along the  $+y$ direction, 
the Lorentz force exerted on the fluctuating element, induces an electric current in the $+z$ direction. Consequently, there appear  fluctuations in density $n$ at the points near the origin on the $z$ axis. This means that in presence of $\textbf{B}=(0,B_y,0)$,  $\pi_x$ is correlated with $n$ with the wave vector
$\textbf{k}=(0,0,k_z)$. As a result, the vector response function $G^R_{\pi_i n}$
can only be made out of $\textbf{V}^3=\textbf{B}\times\textbf{k}$. The point which distinguishes $G^{R}_{\pi_i n}$ from the other retarded response functions
is that the former is not symmetric under the exchange of $\pi_i$ and $n$. It can be simply justified once one applies (\ref{Gab=etaGbaB}) to $G^R_{\pi_i n}$. There are three minus signs which have to be taken into account:
 one from $\eta_{\pi_i}=-1$, one from $\textbf{k}\rightarrow -\textbf{k}$ and the last one from $\textbf{B}\rightarrow -\textbf{B}$. So 
 \begin{equation}
 G^R_{\pi_i n}(\omega, \textbf{k}, \textbf{B})=-G^R_{n \pi_i }(\omega, -\textbf{k}, - \textbf{B})=\,-G^R_{n \pi_i}(\omega, \textbf{k}, \textbf{B})
 \end{equation}
where in the last equality we have used the fact that $ G^R_{\vec{\pi} n}(\omega, \textbf{k}, \textbf{B})=\text{scalar}(\textbf{B}\times \textbf{k})$.

It should be also noted that by taking the limit $\textbf{B}\rightarrow 0$, one can simply reproduce the results of \cite{Kovtun:2012rj}.
\subsection{Kubo Formulas in Presence of a Magnetic Field}
Kubo formulas for transport coefficients of a charged fluid are well known in the literature\cite{Kovtun:2012rj}. However one may be interested in computing transport coefficients for the charged fluid when existing an external magnetic field. Clearly, the response functions are changed compared to $\textbf{B}=0$ case. Therefore the Kubo formulas are expected to be modified too. In table \ref{tabelKubo}, we have listed the relevant  Kubo formulas in presence of a constant magnetic field.   

It is straightforward to check that the $\textbf{B}\rightarrow 0$ limit of formulas given in table (\ref{tabelKubo}) coincide with the Kubo formulas for a charged fluid at zero chemical potential in \cite{Kovtun:2012rj}. Let us emphasis that our Kubo formulas have been expressed in terms of response functions of hydrodynamic densities $n$, $\pi_i$ and $\epsilon$, However, one can also write them in terms of response functions of spacial currents. To proceed, one may first demand the response functions satisfy
\begin{equation}\label{Ward}
\begin{split}
& k^{\mu}\,G_{J_{\mu}J_{\nu}}(\omega. \textbf{k})=\,0\\
& k^{\mu}\,G_{J_{\mu}T_{\rho \sigma}}(\omega. \textbf{k})=\,0\\
& k^{\mu}\,G_{T_{\mu \nu}J_{\sigma}}(\omega. \textbf{k})=\,\eta_{\nu \alpha}F^{\alpha \rho}\,G_{J_{\rho}J_{\sigma}}(\omega. \textbf{k}).
\end{split}
\end{equation}
Then, by use of  (\ref{Ward}) and (\ref{Gab=etaGbaB}), it would be a simple calculation to exchange the response functions of densities with those of spacial currents in Kubo formulas.

\begin{longtable*}{|c|}
	\hline
	\,\,\, Retarded Green's Functions  \\
	\hline
	 \\
	 $\sigma= \lim\limits_{\textbf{k}\rightarrow 0}\, \frac{ - \omega \,\text{Im} G^{R}_{nn}(\omega, \textbf{k})}{ \textbf{k}^2+\,\frac{(\textbf{B}\times \textbf{k})_{i}}{\bar{w}}\,\text{Im} G^{R}_{\pi_{i}n}(\omega, \textbf{k})}
	$\\
	  \\
	\hline
	 \\
	 $\gamma_{\eta}
	 =\,\frac{\mathcal{N}_{\eta}^2}{2 \bar{w}^3 \mathcal{D}_{\eta}}\lim\limits_{\textbf{k}\rightarrow 0}\frac{k^l}{\omega \textbf{k}^2}\partial_{k^l}\left(\mathcal{P}_{i j}^{(1)}\frac{}{}\text{Im}G^R_{\pi_{i}\pi_{j}}\right)-\,\frac{\textbf{B}^2 \,(\textbf{B}\times\hat{\textbf{k}})^2}{ \mathcal{D}_{\eta}}\,\frac{2\sigma^3}{\chi}$
	 \\
	 \\
	\hline
	 \\
	$\gamma_{s}=-\frac{\mathcal{N}_{\eta}^2}{\mathcal{D}_{s}}\lim\limits_{\textbf{k}\rightarrow 0}\frac{\omega k^l}{\textbf{k}^2}\partial_{k^l} \left(\mathcal{P}_{i j}^{(2)}\text{Im}G^R_{\pi_i \pi_j}\right)-\frac{2 (\textbf{B}.\hat{\textbf{k}})^2(\textbf{B}\times\hat{\textbf{k}})^2 \mathcal{D}_{\eta}}{\mathcal{D}_{s}}\bar{w}\sigma^2 \gamma_{\eta}-\frac{4(\textbf{B}\times \hat{\textbf{k}})^2\big((\textbf{B}.\hat{\textbf{k}})^2\textbf{B}^2\sigma^2+\bar{w}^2\omega^2\big)}{\mathcal{D}_{s}}\sigma \bar{w}^2 v_{s}^2$\\
	 \\
	\hline
	\caption{In above expressions we have used $\mathcal{D}_{\eta}=(\textbf{B}^4\sigma^2- \bar{w}^2 \omega^2)$, $\mathcal{N}_{\eta}=(\textbf{B}^4\sigma^2+ \bar{w}^2 \omega^2)$,
		$\mathcal{D}_{s}=\mathcal{N}_{\eta}\big(\mathcal{N}_{\eta}-2 (\textbf{B}\times\hat{\textbf{k}})^4 \bar{w}^2 \sigma^2 \omega^2\big)$.
		\newline
		 The projective tensors are defined as:
		\newline
		$\mathcal{P}_{i j}^{(1)}=\delta_{i j}-\csc^2\theta\,\hat{k}_{i}\hat{k}_{j}- \csc^2 \theta \hat{B}_{i} \hat{B}_{j}+ \cos \theta \csc^2 \theta\, \hat{k}_{(i} \hat{B}_{j)}$ 
		\newline
		 $\mathcal{P}_{i j}^{(2)}=\cot^2\theta\,\hat{k}_{i}\hat{k}_{j}+\csc^2 \theta \hat{B}_{i} \hat{B}_{j}- \cos \theta \csc^2 \theta\, \hat{k}_{(i} \hat{B}_{j)}$}\label{tabelKubo}
\end{longtable*}
\section{Outlook}
Our results reported in this note might be encountered with a very important
follow-up question. The question is how the hydrodynamic modes would be affected if we took into account the chiral effects in the fluid. Very recently, Yamamoto has shown that in a system with gravitational anomaly, there exists a new type of collective mode. As shown
in  \cite{Yamamoto:2015ria}, the so-called Chiral Alfven wave is a transverse hydrodynamic mode propagating in an incompressible fluid, with a velocity proportional to the anomaly coefficient. On the other hand, analogous to what we have done in this paper, one can study the dissipative effect in a chiral fluid. Specifically, it would be interesting to investigate whether the Alfven mode splits into two or more chiral modes when taking into account the dissipative effects or not. 

In another direction one may study the hydrodynamic modes and response functions in a charged Lifshitz fluid \cite{Roychowdhury:2015jha,Hoyos:2013qna}. We leave more investigation on the issue to our future works.

$Acknowledgements$: We would like to thank to Pavel Kovtun for exchanging three emails. We would also like to thank to Piotr Surowka for exchanging one email.

\bibliographystyle{utphys}


\providecommand{\href}[2]{#2}\begingroup\raggedright\endgroup

\end{document}